\documentclass{article}
\usepackage{amsmath}
\usepackage{color}
\definecolor{blue}{rgb}{0.0, 0.0, 1.0}
\definecolor{red}{rgb}{1.0, 0.0, 0.25}
\definecolor{green}{rgb}{0.01, 0.75, 0.24}

\usepackage{multirow}% http://ctan.org/pkg/multirow
\usepackage{hhline}% http://ctan.org/pkg/hhline
\usepackage{cite}

\newcommand{\cmmnt}[1]{\ignorespaces} 

\usepackage{url}
\usepackage{todonotes}
\usepackage{tabularx}
\usepackage{array}
\usepackage{graphicx} 
\usepackage{multirow}

\usepackage[ruled]{algorithm2e}
\usepackage{multirow}

\usepackage{hhline,booktabs,amsmath}
\usepackage{subcaption}

\usepackage{arxiv}
\usepackage[utf8]{inputenc} % allow utf-8 input
\usepackage[T1]{fontenc}    % use 8-bit T1 fonts
\usepackage{hyperref}       % hyperlinks
\usepackage{url}            % simple URL typesetting
\usepackage{booktabs}       % professional-quality tables
\usepackage{amsfonts}       % blackboard math symbols
\usepackage{nicefrac}       % compact symbols for 1/2, etc.
\usepackage{microtype}      % microtypography
\usepackage{lipsum}

\title{QuPWM:  Feature Extraction Method for MEG Epileptic Spike Detection }

\author{\\
Abderrazak Chahid$^{ 1}$,  Fahad~Albalawi$^{ 1,2}$,  Turky Nayef Alotaiby$^{3}$, Majed Hamad Al-Hameed$^{4}$,\\ Saleh Alshebeili$^{5, \star}$, Taous-Meriem Laleg-Kirati $^{ 1, \star}$
\thanks{$^{1}$ Computer, Electrical and Mathematical Sciences and Engineering Division (CEMSE) King Abdullah University of Science and Technology (KAUST), KSA. E-mail:abderrazak.chahid@kaust.edu.sa, taousmeriem.laleg@kaust.edu.sa }% <-this % stops a space
\thanks{$^{2}$ Electrical Engineering Department, Taif University, Taif, 21955, KSA.  E-mail: fahad.albalawi@kaust.edu.sa  }
\thanks{$^{3}$King Abdulaziz City for Science and Technology (KACST), Riyadh , KSA. E-mail: totaiby@kacst.edu.sa }% <-this % stops a space
\thanks{$^{4}$ National Institute of Neuroscience, King Fahad Medical City, Riyadh 11525, KSA. }% <-this % stops a space
\thanks{$^{5}$ King Saud University (KSU), Riyadh, KSA. E-mail:dsaleh@ksu.edu.sa }% <-this % stops a space
\thanks{$^{\star}$ Corresponding authors e-mails: taousmeriem.laleg@kaust.edu.sa, dsaleh@ksu.edu.sa}}

\begin{document}

% keywords can be removed

\maketitle

\begin{abstract}
Epilepsy is a neurological disorder  classified as the second most serious neurological disease known to humanity, after stroke. Localization of epileptogenic zone is an important step for  epileptic patient treatment, which starts with epileptic spike detection. The common practice for spike detection of brain signals  is via visual scanning of the recordings, which is a subjective and a very time-consuming task. Motivated by that, this paper focuses on using machine learning for  automatic detection of epileptic spikes in magnetoencephalography (MEG) signals. First, we used the Position Weight Matrix (PWM) method  combined with a uniform quantizer  to generate useful features. Second, the extracted features are classified using  a Support Vector Machine (SVM) for the purpose of epileptic spikes detection. The proposed technique shows great potential in improving the spike detection accuracy and reducing the feature vector size. Specifically, the proposed technique achieved average accuracy  up to  98\% in  using  5-folds cross-validation applied to a balanced dataset of 3104 samples. These  samples are  extracted from 16 subjects where eight are  healthy and eight are epileptic subjects using a sliding frame of size of 100 samples-points with a step-size of 2 sample-points.
\end{abstract}


% \listoftodos

%\IEEEpeerreviewmaketitle

\section{Introduction}

  Epilepsy is not a singular disease entity but a variety of dysfunctions reflecting brain disorders or abnormal electrical activities that may strike patients of all ages \cite{Epidef}. An epileptic seizure occurs when a burst of electrical impulses in the brain exceeds its normal limits. These pulses spread to neighboring areas in the brain, which may create an uncontrolled storm of electrical activity sent to body organs. The electrical impulses could be transmitted to the muscles, causing twitches or convulsions. In particular, epileptic patients may stare blankly for a few seconds during a seizure, while others have uncontrollable jerking movements of the arms and legs. Among the diagnosis tools for epilepsy is the magnetoencephalography (MEG) \cite{Hamalainen1993}. The MEG is a recent functional neuro-imaging technology that measures the magnetic activity of the brain. This new technology uses an array of highly sensitive sensors or magnetometers called superconducting quantum interference devices (SQUIDs). However, as the brain magnetic field is very weak compared to surrounding magnetic sources, the MEG measurement needs a shielded room to isolate the patient from the external magnetic fields such as the magnetic field of the earth or the electronic devices. For Epileptic patients, the MEG is used during two main phases of the treatment \cite{Stefan2017a,Englot2015}: first, localizing the region of the brain which produces the abnormal electrical activities that cause the neurological disorder. If the localized region is not responsible for a vital function in the brain such as speech, this region is usually extracted surgically if the medications were not effective for the treatment. Second, assessing surgeries outcomes.

Neurologists often spend hours to manually read the MEG recordings. For this reason, automated interictal epileptic spike detection in MEG signals has attracted research interest over the last decade.  Four methods, to the best of our knowledge,  have been proposed for spike detection of the MEG signals in literature \cite{Baillet2017,El-Samie2018}. The first method uses independent component analysis (ICA).The ICA method  is a multi-channel MEG spikes localization method that decomposes spike-like and background components into separate spatial topographies and associated time series \cite{Ossadtchi2004}. The detection is performed using a thresholding technique. The second method is based on the common spatial patterns (CSP) method and linear discriminant analysis (LDA) (CSP-LDA) \cite{Khalid2016}. The CSP-LDA performs eigenvalue decomposition of the covariance matrices of the input data to find the most discriminative features. LDA classifier is then employed to perform the spikes detection. The third method is the Amplitude Thresholding and Dynamic Time Warping (AT-DTW) proposed in \cite{Khalid2017}. This method first uses amplitude thresholding to determine the most likely spiky segments. For the spike detection, the method employs dynamic time warping. The maximum specificity and sensitivity reported in the aforementioned detection methods are 95.8\%,and 92.4\%, respectively.   A more recent detection method has been proposed in \cite{chah2019SCSA} where the decomposition of the signal into squared eigenfunctions of the Schrodinger operator was used  for detection\cite{Laleg-Kirati2013}.  This method uses the largest negative eigenfunction in absolute value  of  the discrete spectrum of the  Schr\"{o}dinger Operator  as a feature with the   Random Forest (RF) classifier. With this low dimension feature vector this method could achieve a sensitivity and specificity of  93.68\% and 95.08\%, respectively. 

  \begin{figure*}[!t]    
\centering
   \includegraphics[width=0.92\linewidth]{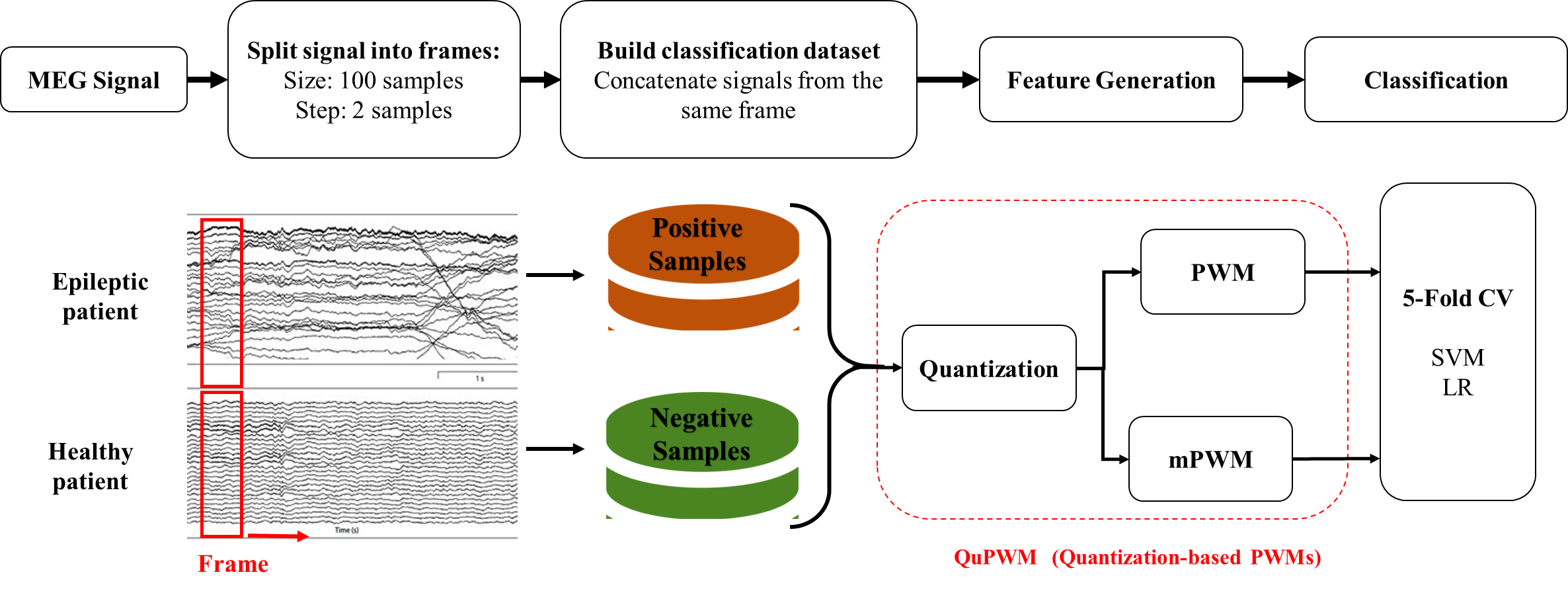}
    \caption{Classification framework subdivided into three stages: MEG records pre-processing, feature generation and classification models. }
    \label{fig:Frame_Work_MEG}
\end{figure*}

 In this work, we propose a novel feature generation method for multi-channel MEG signal which improves patient-independent interictal spike detection. This method is based on the combination of the Position Weight Matrix (PWM) method and uniform quantization scheme. We name this approach QuPWM.  This method  takes advantage of the efficiency of the PWM method, which is usually used for DNA sequences classification. QuPWM shows great potential in improving the accuracy of the interictal spike detection models.

The paper is organized as follows. Section II  includes a description of the MEG dataset and the proposed features extraction process methods based on the PWM method and the used quantization schemes in Section II. Section III presents the obtained   results   using eight healthy and eight epileptic patients which  are presented in Section III. Section IV  presents a discussion of the findings, where  a conducted sensitivity analysis on the  proposed feature extraction method is analyzed with respect to the frame length and number of subjects. Finally, Sections V summarizes our concluding remarks.
  
  \begin{figure}[!h]    
\centering
   \includegraphics[width=0.75\linewidth]{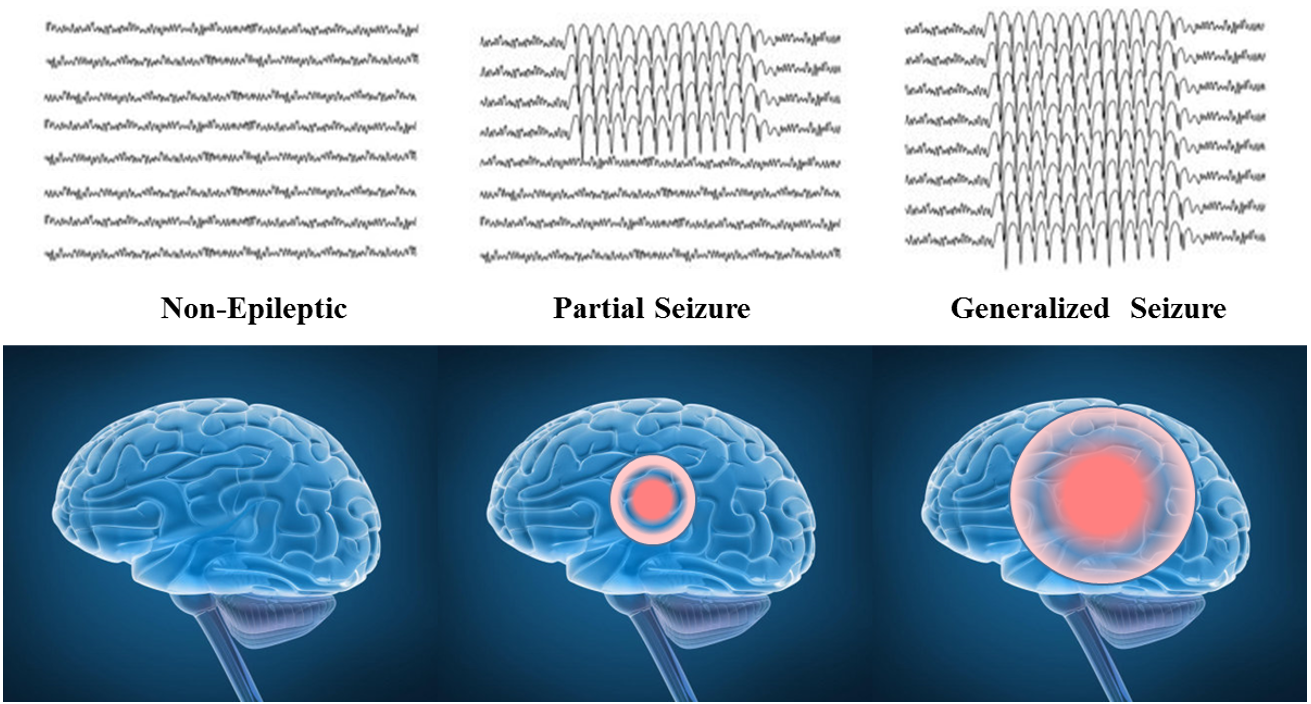}
    \caption{Illustration of the brain abnormal activities in different types of epileptic seizures [built based on  \cite{brain_2018} \cite{bates_2018}].}
    \label{fig:Epeliptice_types}
\end{figure} 

\subsection{Classification problem definition}
An Epileptic seizure happens when a burst of electrical impulses in the brain exceed  their normal limits. They spread to neighboring areas in the brain and might create an uncontrolled storm of electrical activity sent to body organs. The electrical impulses can be transmitted to the muscles, causing twitches or convulsions.

For epileptic patient treatment, it is important  to localize the region in the brain that has an abnormal activity to extract this part with surgical intervention. After the surgery, the patient needs to take MEG test to ensure that the abnormal brain activities are absent. Up to now, the visual assessment is the "only" way to assess the MEG signals. Indeed, there is  no efficient  computer based aid to analyse MEG signals. Therefore, there is a need for developing machine learning based algorithms to assist clinicians in quickly and efficiently detecting and predicting  abnormalities in MEG signals. 
% In this study, the epileptic spikes detection problem aims to distinguish between spiky and healthy MEG signals recorded using 24 electrodes. 

\section{ Materials and Methods} 

The proposed methodology consists of three main  steps as shown in Figure \ref{fig:Frame_Work_MEG}.  First, the MEG recordings are split into segments using framing technique. For instance, the first frames of the 24 channels are concatenated to form the first sample, and so on.  Then  the samples are converted into sequences using a uniform quantization scheme. Second, the quantified samples are mapped to extract two types of PWM-based features. Finally, the classification model SVM is used for spike detection. These three steps are explained in the following sections \ref{MEG_prepro} and \ref{Feature_GENr}.

\subsection{Experimental data acquisition and analysis}\label{Datata_frame}

MEG data were recorded in a shielded room at National Neural Institute- King Fahad Medical City (NNI-KFMC), Riyadh, Saudi Arabia with an Elekta Neuromag system. As the MEG signals are much weaker than normal environmental magnetic noise, the shielded room blocks the majority of environmental magnetic fields so that the magnetic fields generated by the brain can be accurately  detected. Elekta Neuromag head system (helmet) contains 102 magnetometer and 204 gradiometer sensors. These sensors are further categorized according to the different brain regions. Clinically, the brain is divided into eight regions; left temporal (LT), right temporal (RT), left frontal (LF), right frontal (RF), left parietal (LP), right parietal (RP), left occipital (LO), and right occipital (RO). Each element of the Elekta Neuromag system is comprised of three sensors, one magnetometer, and two gradiometers. Magnetic brain activity was recorded at a sampling frequency of 1 kHz. MEG data was filtered by tSSS (Spatiotemporal signal space separation) method \cite{Taulu_2006}. The data were then off-line band-pass filtered 1–50 Hz for visual inspection. A total of 18 MEG data segments, each of 15 minutes duration and 26 channels, were taken from 8 epileptic patients and eight healthy patients. These segments are analyzed by specialized  neurologists from NNI, KFMC, Riyadh. The neurologists marked the MEG spikes locations, in different brain regions, by visual inspection. The total number of spikes in these recordings is 166. As mentioned earlier, there are 306 sensors to cover the whole head. These sensors are further marked according to the brain regions.

 Written informed consent was signed by each participant or responsible adult before they participated in the study. The study was conducted in accordance with the approval of the Institutional Review Board at KFMC (IRB log number: 15-086, 2015).

 \begin{figure} [!h]   
\centering
   \includegraphics[width=0.75\linewidth]{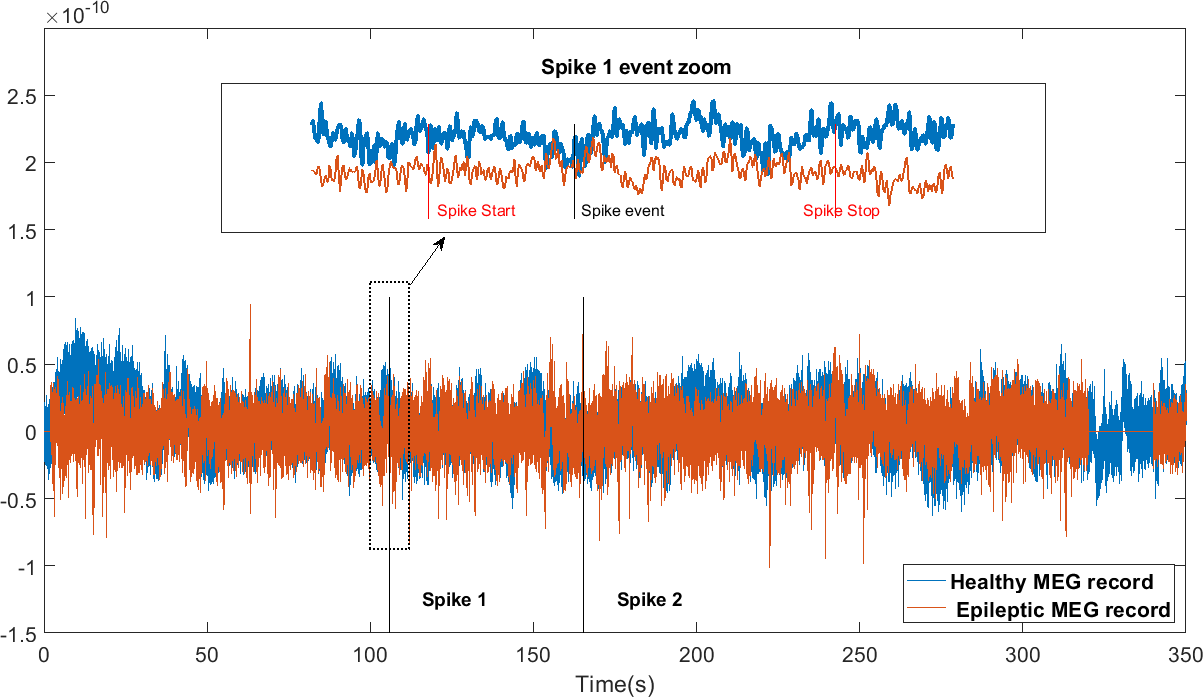}
    \caption{Example of healthy and epileptic subject MEG recording. Zoomed plot shows a segment where an epileptic spike event happens in epileptic subject (orange) compared to a healthy subject (blue). }
    \label{fig:MEG_2a}
\end{figure}

\subsection{Multi-channels MEG signal pre-processing}\label{MEG_prepro} 

 Multi-channels MEG  pre-processing  aims essentially to prepare classification samples.   The pre-processing consists of two steps: 
\begin{itemize}
\item \textbf{\textit{signal Framing}}:   build equisized samples that  combine  sliding widows or frames from all  the 24 MEG channels. These combined frames will represent the raw data samples used as inputs to the  classification which will be quantified in the next step. 
\item  \textbf{\textit{signal Quantization}}: convert the real-values samples to sequences of letters. This step is needed to make the samples suitable for the  used feature generation method explained in Sections  \ref{FeatEXTR} and \ref{mFeatEXTR}.
\end{itemize} 

\vspace{1cm}

\subsubsection{MEG signal Framing}
The multi-channel MEG signals are segmented into frames using sliding frames using a sliding window of size $100$ sample-points with  a step of $2$ sample-points as shown in Figures \ref{fig:Frame_Work_MEG} and \ref{fig:MEG_2a}. For instance,    the first  100 sample-points of all  channels are concatenated together to form  the first sample of size 2400 sample-points  of the input raw-data. A binary label or class is assigned to each sample as follows:

\begin{itemize}
\item \textbf{\textit{Positive class}}: represent samples that have epileptic spike event occurring in MEG records of the epileptic patients.
\item  \textbf{\textit{Negative class}}: represent samples from uniformly distributed time-locations in MEG records of the healthy patients.
\end{itemize}

After signal framing, the positive and negative samples will be quantified using a uniform quantizer as explained in the next section.  

\vspace{0.5cm}

\subsubsection{ MEG signal quantization}\label{Quant}
\begin{figure}[!h]  
\centering

   \includegraphics[width=0.75\linewidth]{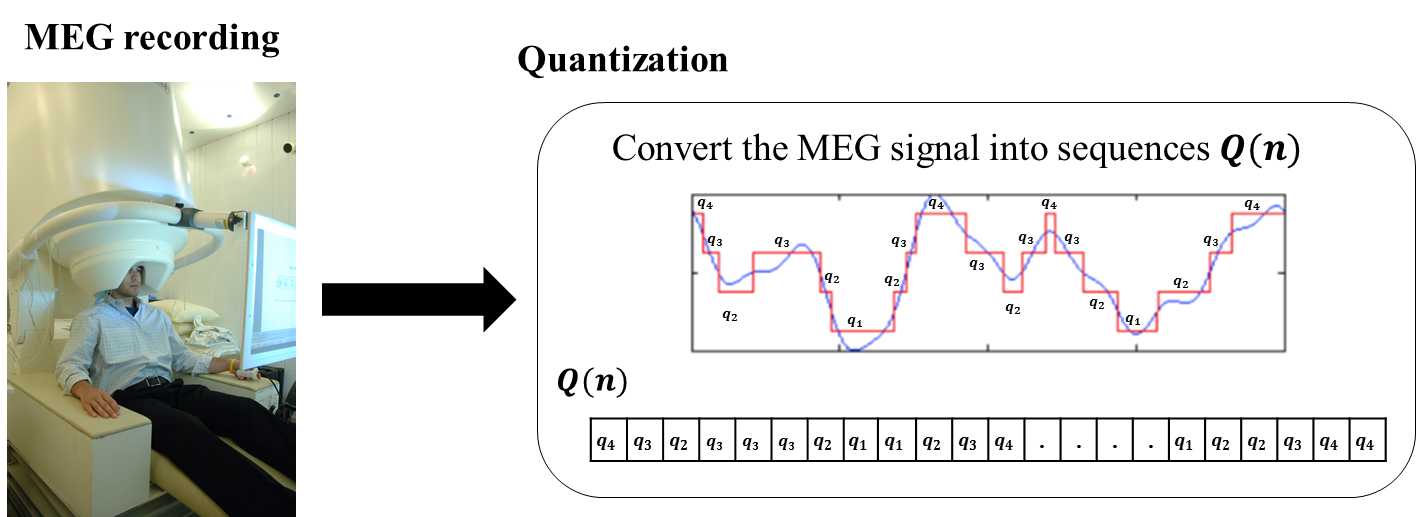}
    \caption{Example of MEG signal quantization using four levels.   }
    \label{fig:Quantization}
\end{figure}

In this work we employed the PWM method for feature extraction as explained in Section \ref{FeatEXTR}. However, as the PWM method deals only with sequences, the input MEG samples should be converted into sequences, as shown in Figure \ref{fig:Quantization}. For this reason, we used a uniform quantization scheme.  The quantization  is utilized to convert the real-valued signal  $X$ to a sequence $Q$ of different  levels $q_1, q_2, ..., q_M$ defined as follows \cite{gray1998quantization,pollard1982quantization}:

\begin{equation}\label{Quants}
Q(n)  =  q_i ~~~~~  \mbox{ where }  ~~~~~ i=1,2,\dots ,M
\end{equation}

One of the purposes of using quantization in this classification is to illuminate noise effect by mapping a range of value of  the input  signal,  as presented in Figure  \ref{fig:Quantization}, to a single level $q_i$  based on the probability density of this range of values in the dataset. The quantization scheme depends on two major parameters which are the desired number of levels  $M$ and the resolution $\mathbf{r}$. For an appropriate choice of these parameters, the probability distribution of the  signal is analyzed for both classes. 

\begin{figure}[!h]    
\centering

   \includegraphics[width=0.75\linewidth]{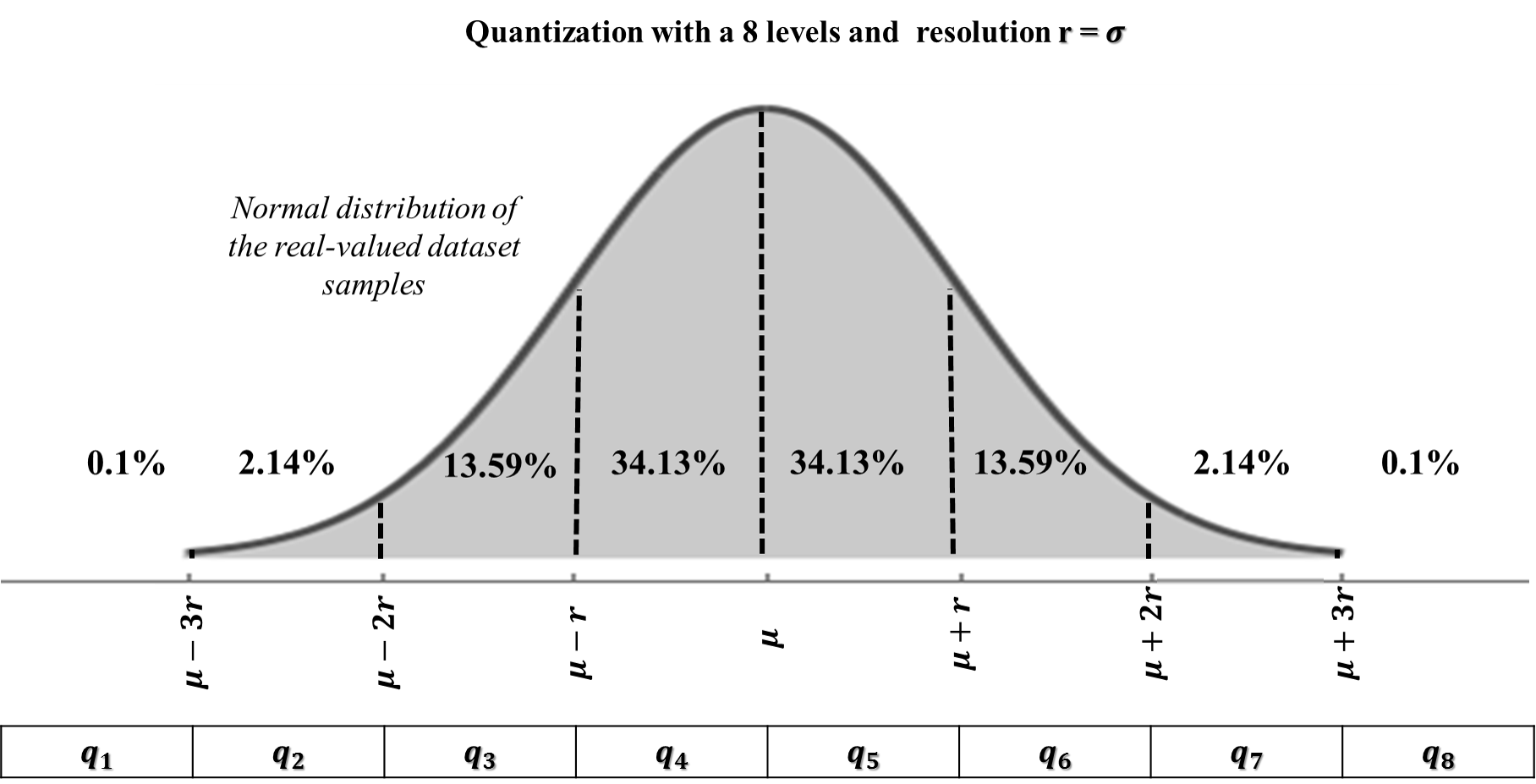}
    \caption{The quantization of the real-valued sequence with a resolution $r=\sigma$ and $M=8$. The percentage values reflect the significance of probability of each  interval.}
    \label{fig:Level_resolution}
\end{figure}

Indeed, it is known in  probability theory and mathematical statistics that the value of a random variable is considered impossible, in other words abnormal, if it lies  $3\sigma$ far from its mean $\mu$. This observation is called the 3-sigma rule \cite{nikulin2011three} \cite{3Sigma2011}.   Therefore, the most significant variability and randomness that characterizes a Gaussian random variable $X$, where the other value can be considered as outlier values which can be neglected. This fact is observed if : 
\begin{equation}
    \vert X(n)-{\mu}\vert \leq 3{\sigma}
\end{equation}
where $\mu$ and $\sigma$ are defined to be  the mean and  standard deviation of  the Gaussian random variable $X$. \\

The standard rule is used to define quantization scheme as shown in  Figure \ref{fig:Level_resolution}. It is important to note that the probability distribution of  $X$ varies from subject to subject which affects the quantization scheme. Therefore and in order to unify the quantization scheme, the mean $\mu$ and  standard deviation $\sigma$ are defined  as the average mean and  standard deviation of all  subjects as follows:

\begin{equation}\label{sigma0}
\mu  =   \dfrac{1}{N_s} \displaystyle \sum_{n=1}^{N_s}  \mu_{n}  ~~~\mbox{and}~~~  \sigma  =   \dfrac{1}{N_s} \displaystyle \sum_{n=1}^{N_s}  \sigma_{n}~,
\end{equation}

\noindent where $N_s$ is the number of subjects. $\mu_{n}$ and $\sigma_{n}$ are the mean and the standard deviation of the data values of the $n^{th}$ subject.\\

Since the 3-sigma rule indicates that most of the information of a Gaussian random variable are located within $3\sigma+\mu$, the quantization resolution is set to be equal to the  standard deviation $\sigma$ as shown in Figure \ref{fig:Level_resolution} and explained in detail in Algorithm 1.%\ref{ererer} . 

\begin{algorithm}
\label{ererer}
\caption{ Quantization Algorithm}

          \textbf{Input :} $X$: Real-valued MEG signal \\
             $~~~~~~~~~M$: Number of Quantization levels \\
             $~~~~~~~~~\mu$:  Quantization centroide \\
             $~~~~~~~~~\mathbf{r}$:  Quantization resolution \\
\vspace{0.1cm}

          \textbf{Output:} $Q$: Output  sequence\\ 
\vspace{0.3cm}

%  \diamond ~~  \textit{ Generate the $M$ levels set }  \\
 $N_X$ =  \mbox{length of} $~~X$
\vspace{0.3cm}

\For{$n\gets1$ \KwTo $N_X~~~$}{
\vspace{0.5cm}

    \For{$k\gets \dfrac{2-M}{2} $ \KwTo $\dfrac{M}{2} ~~~$}{
\vspace{0.5cm}

          \uIf{$   X(n) < \mu + \dfrac{(2-M)~\mathbf{r}}{2} ~~~ $}{
                  \vspace{0.2cm}
                    $Q(n)=q_{1}$;
                   \vspace{0.5cm}

          }

          \uElseIf{ $\mu + (k-1)~ \mathbf{r}  \leq  X(n) < \mu + k~ \mathbf{r}~~~ $}{
                  \vspace{0.2cm}

                    $Q(n)=q_{(k+M/2)}$;
                   \vspace{0.5cm}

          }
            \uElse{
                    $Q(n)=q_{M}$;

          }
     }
    \vspace{0.5cm}

}    
\end{algorithm}

 \subsection{QuPWM-based features generation }\label{Feature_GENr}

The proposed Quantization-based   position weight matrix (QuPWM)  feature generation method is based on combining the position weight matrix (PWM) method with Quantization.  This method uses two approaches to generate the PWM features  as explained in the next section. 
\vspace{0.5cm}

\subsubsection{ Standard PWM-based features }\label{FeatEXTR}

A position weight matrix (PWM) , also known as a position-specific weight matrix (PSWM) or position-specific scoring matrix (PSSM), has been presented for the first time by Gary  in \cite{stormo1982use}. This method  is  widely used technique for  motifs  characterization and discovery in biological sequences such as DNA/mRNA \cite{hertz1999identifying,staden1984computer,Akhtar2010,Tabaska1999,xia2012position}. It showed high potential in sequences characterization and motifs  extracting with remarkable binary classification accuracy. For MEG signals, we  adapted the  PWM-based method to extract relevant features after signals quantization methods. The PWM is based on building two matrices, $PWM^+$ derived from the positive training set and the second $PWM^-$ represents the negative set. The PWM matrices indicate the significance of each position along the input sequences for each class.  For binary classification, the two matrices  $PWM^+$  and $PWM^-$ are defined as follows: 

\begin{equation}\label{PWMss}
PWM^+[i,q_j ] = \sum_{s=1}^{N^+_s} \xi (Q^+_s(i),q_j) 
\end{equation}
and 
\begin{equation}\label{PWMss}
PWM^-[i,q_j] = \sum_{s=1}^{N^-_s} \xi (Q^-_s(i),q_j)
\end{equation}

where $N^+_s$ and $N^-_s$ are  the total number of sequences in the positive and negative classes.  The $\xi(a,b)$ function is defined as follows:

\begin{equation}\label{xi_func} 
\xi(a,b)= \begin{cases} 
1 & ~  \mbox{if the $a=b$ }\\ 
0  &  ~  elsewhere
\end{cases} 
\end{equation}

 Table \ref{tab:PWM_example} shows an example of a positive matrix  $PWM^+$. Each column of the matrix  represents the frequency  of a specific letter $q_i$ in a specific position $n$  among all positive sequences.

 \begin{figure*}[t]   
\centering
  \includegraphics[width=0.95\linewidth]{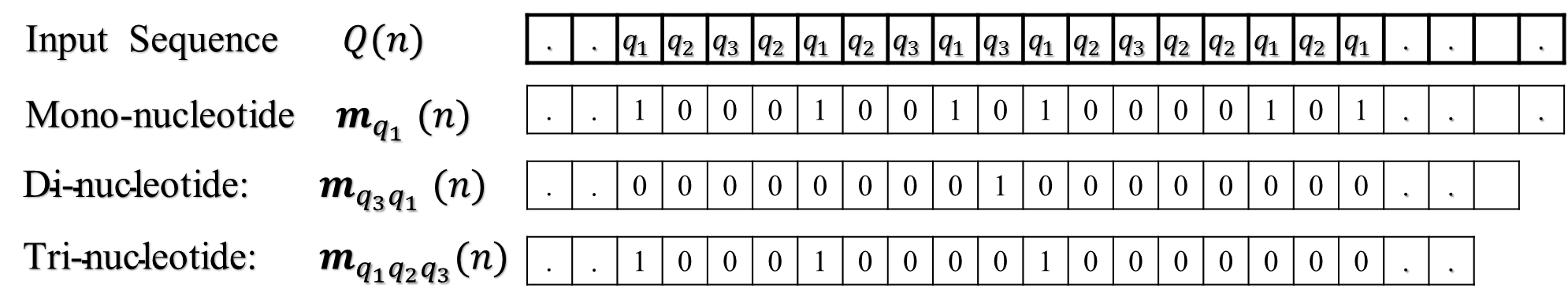}
    \caption{ Examples of three binary sequences corresponding to mono-mers, di-mers and tri-mers, respectively. }
    \label{fig:Pattern_ACGT}
\end{figure*}

 \begin{table}[!h]
 \centering

      \caption{Example of a position-wight matrix $PWM^+$ containing the frequencies of five levels $[q_1, q_2, \dots, q_5]$. }
    \label{tab:PWM_example}
    
   \resizebox{0.5\linewidth}{!}{

   \renewcommand{\arraystretch}{1.3}

\begin{tabular}{cccccc}
\hline
\multirow{2}{*}{\begin{tabular}[c]{@{}c@{}}$n^{th}$ \\ position\end{tabular}}  & \multicolumn{5}{c}{\textbf{Quantization levels}} \\ \cline{2-6} 
& \textbf{$q_1$} & \textbf{$q_2$} & \textbf{$q_3$} & \textbf{$q_4$} & \textbf{$q_5$} \\ \hline

1 & 1104 & 257 & 331 & 384 & 372 \\
2 & 1209 & 213 & 271 & 291 & 324 \\
3 & 1217 & 245 & 298 & 336 & 336 \\
4 & 987 & 263 & 309 & 409 & 474 \\
5 & 941 & 253 & 383 & 442 & 532 \\
6 & 1273 & 243 & 308 & 317 & 367 \\
7 & 1494 & 208 & 243 & 261 & 245 \\
8 & 1301 & 264 & 299 & 364 & 388 \\
9 & 61 & 101 & 306 & 829 & 1331 \\
10 & 144 & 88 & 266 & 599 & 1133 \\
% 11 & 145 & 163 & 385 & 730 & 1150 \\
% 12 & 1212 & 226 & 289 & 304 & 312 \\
% 13 & 1166 & 224 & 281 & 322 & 383 \\
% 14 & 1239 & 204 & 264 & 368 & 405 \\
% 15 & 1192 & 248 & 297 & 350 & 412 \\
% 16 & 1440 & 219 & 255 & 295 & 269 \\
.. & .. & .. & .. & .. & .. \\ \hline
\end{tabular}

   }
\end{table}

Then, these two PWMs are used to generate two  scores representing the extracted features. The two  scores $Score^+(\textbf{Q})$ and  $Score^-(\textbf{Q})$ represent the two probabilities of a given  sequence to be in the positive or negative class. In other words, for a given sequence $Q$ to be positive, the score induced by $PWM^+$ should be greater than the score induced by $PWM^-$. The  two scores $Score^+(\textbf{Q})$ and  $Score^-(\textbf{Q})$  of a sequence $Q$  are defined as as follows: 

\begin{equation}\label{fPWM}
Score^+(\textbf{Q}) = \sum_{n=1}^{N} \frac{PWM^+ \left[  n,Q(n) \right]}{\sum_{j=1}^{M} PWM^+[n,q_j]}
\end{equation}

and 

\begin{equation}\label{fPWM}
Score^-(\textbf{Q})  = \sum_{n=1}^{N} \frac{PWM^- \left[  n,Q(n) \right]}{\sum_{j=1}^{M} PWM^-[n,q_j]}
\end{equation}

where $N$ is the number of sample-points in the sample $Q$.

It is very important to mention that the PWMs should be reconstructed only from the training dataset  in order not to violate the classification rules.

\vspace{0.5cm}

\subsubsection{ Motif-based PWM features}\label{mFeatEXTR}

In order to extract more advanced patterns, we adopted a new approach which  deals with binary sequences, extracted from the original sequence,  that represent the presence of a specific motif of one letter or more in the original sequence, as shown in Figure  \ref{fig:Pattern_ACGT}.  We called this approach  the motif-based Position Weight Matrix (mPWM). The main idea is to decompose every sequence into binary sequences reflecting a specific pattern of levels in this sequence  such as :mono-mers ($q_1$, $q_2$, $q_3$, etc), di-mers ($q_1q_1$ , $q_1q_2$ , $q_1q_3$, etc), tri-mers ($q_1q_1q_1$, $q_1q_2q_3$, $q_3q_3q_3$, etc). For instance, a sequence  of  $3$ levels $q_1$, $q_2$, $q_3$ will give: $3$ possible mono-mers, $3x3 = 9$ possible di-mers combinations, and $3x3x3 = 27$ possible tri-mers combinations. Figure  \ref{fig:Pattern_ACGT} shows the different \textit{k}-mer encodings. For instance, the binary sequence of the mono-mers $m_{q_1}$ is defined as follows:

\begin{equation}\label{mapping2} 
m_{q_1}(n)= \begin{cases} 
1 & ~  \mbox{if  $Q(n)=q_1$  }\\ 
0  &  ~  elsewhere
\end{cases} 
\end{equation}

Similarly, the binary sequence of the di-mers $m_{q_1q_3}$ is defined as follows:
\begin{equation}\label{mapping2} 
m_{q_1q_3}(n)= \begin{cases} 
1 & ~  \mbox{if  $Q(n)=q_1$ and  $Q(n+1)=q_3$  }\\ 
0  &  ~  elsewhere
\end{cases} 
\end{equation}

The binary mapping of  \textit{k}-mers motif extraction is summarized in Algorithm 2.

Similarly to the standard PWM,  the binary sequences $m_j$, representing the     \textit{k}-mer  motifs  $\textbf{j} \in \Omega_k $, which is defined in Eq. \ref{XC}, is used to reconstruct multiple pair of  PWM  matrices  $PWM_j^+$  and $PWM_j^-$ defined as follows: 

\begin{equation}\label{PWMss}
PWM_\textbf{j}^+(n) = \sum_{s=1}^{N^+_s} \xi (m_{j,s}^+(n),1) 
\end{equation}
and

\begin{equation}\label{PWMss}
PWM_\textbf{j}^-(n) = \sum_{s=1}^{N^-_s} \xi (m_{j,s}^-(n),0)
\end{equation}

where  $\textbf{j} \in \Omega_k $, such that $\Omega_k$ represents the set  of the possible  \textit{k}-mers combination defined as:
\begin{equation}\label{XC}
\begin{cases} 
\Omega_1 =\{q_1, ~q_2, ~q_3,\dots\} ~~~~~~~~~~~~~~~~~~\mbox{'mono-mers'}\\
\Omega_2 =\{q_1q_1, ~q_1q_2, ~q_1q_3,\dots\}~~~~~~~~~~~ \mbox{'di-mers'}\\
\Omega_3 =\{q_1q_1q_1, ~q_1q_2q_3, ~q_3q_3q_3,\dots\} ~~~~\mbox{'tri-mers'}\\
\dots\\
\end{cases} 
\end{equation}

 $m_{j,s}$ denotes the motif $m_{j}$ extracted from the $s^th$ sequence $Q_s$ .  $N^+_s$ and $N^-_s$ are  the total number of sequences in the positive and negative classes.  

\begin{algorithm}[!h] \label{algo-kmer}
\caption{\textit{k}-mers  motif extraction}

  \textbf{Input~~} Q: the input sequence, k: the order of  \textit{k}-mers\\ 
  \textbf{Output~} \textit{k}-mers: the motifs   of \textit{k}-mers   \\
% \Procedure{\textit{k}-mers=-merss\_patterns\_extraction}{$Seq$, $d$}
---
\vspace{0.2cm}

 $i\gets 1$ \\
 $N\gets $ size(Q) \\
 $C_d\gets$ {all possible combinations composed of \textit{k}-mers}\\
\vspace{0.3cm}

\For{each combinations}{
\vspace{0.3cm}

 $C_d(i)\gets$ new combination 
\vspace{0.3cm}

\For{$j=1:N-d+1$}{
\vspace{0.3cm}

  \uIf{$Q(j:j+d-1)= C_d(i)$ }{
    \textit{k}-mers$(i,j) \gets 1 $}
  \uElse{
    \textit{k}-mers$(i,j) \gets 0 $ \\
  }
 }

 $i\gets i+1$ 
}
%  \Return \textit{k}-mers
\end{algorithm}

%   \begin{figure}[!h]
%       \centerline{\includegraphics[width=\linewidth]{.PWM_frame}}
%       \caption{Illustration of  \textit{k}-mer motifs for a sequence of letters \{ A,B,C,D,....\} }
%       \label{fig:PWM_frame_kmer}
%       \centering
%   \end{figure}

%  \begin{figure}[!h]
%       \centerline{\includegraphics[width=\linewidth]{.PWM8_example}}
%       \caption{Example of pattern-based position weight matrix $PWM^+$ for  \textit{k}-mers. }
%       \label{fig:PWM8_example}
%       \centering
%   \end{figure}

\begin{table}[!t]

 \centering

      \caption{Example of  position-wight matrix $PWM_{\textbf{j}}^+$ containing the frequencies of di-mer motif. }
    \label{tab:m2PWM_example}
    
   \resizebox{0.6\linewidth}{!}{
   \renewcommand{\arraystretch}{1.7}

\begin{tabular}{cccccccccc}
\hline
\multirow{2}{*}{\begin{tabular}[c]{@{}c@{}}$n^{th}$ \\ position\end{tabular}} & \multicolumn{9}{c}{\textbf{The di-mer motifs (j) }} \\ \cline{2-10} 
 & \textbf{$q_1q_1$} & \textbf{$q_1q_2$} & \textbf{$q_1q_3$} & $q_1q_4$ & \textbf{$q_2q_1$} & \textbf{$q_2q_2$} & \textbf{$q_2q_3$} & \textbf{$q_2q_4$} & \textbf{..} \\ \hline
1 & 15 & 13 & 11 & 14 & 17 & 13 & 11 & 8 & .. \\
2 & 10 & 10 & 12 & 5 & 9 & 11 & 16 & 6 & .. \\
3 & 0 & 0 & 0 & 0 & 0 & 0 & 0 & 0 & .. \\
4 & 0 & 0 & 0 & 0 & 0 & 0 & 0 & 0 & .. \\
5 & 8 & 10 & 8 & 12 & 7 & 14 & 3 & 12 & .. \\
6 & 298 & 290 & 310 & 294 & 303 & 293 & 322 & 299 & .. \\
7 & 93 & 93 & 101 & 112 & 94 & 109 & 89 & 116 & .. \\
8 & 0 & 0 & 0 & 0 & 0 & 0 & 0 & 0 & .. \\
% 9 & 0 & 0 & 0 & 0 & 0 & 0 & 0 & 0 & .. \\
% 10 & 85 & 119 & 96 & 105 & 104 & 110 & 89 & 110 & .. \\
% 11 & 724 & 694 & 690 & 686 & 699 & 683 & 704 & 680 & .. \\
% 12 & 3 & 5 & 6 & 3 & 1 & 5 & 1 & 7 & .. \\
% 13 & 0 & 0 & 0 & 0 & 0 & 0 & 0 & 0 & .. \\
% 14 & 0 & 0 & 0 & 0 & 0 & 0 & 0 & 0 & .. \\
% 15 & 1 & 5 & 3 & 6 & 5 & 2 & 4 & 2 & .. \\
% 16 & 4 & 2 & 4 & 4 & 2 & 1 & 2 & 1 & .. \\
.. & .. & .. & .. & .. & .. & .. & .. & .. & ..\\ \hline
\end{tabular}

}
\end{table}

 Similarly, the  two scores $Score^+(m_j)$ and  $Score^-(m_j)$  of a  \textit{k}-mer $m_j$,  extracted from a sequence $Q$, are defined as as follows: 

\begin{equation}\label{mfPWMp}
Score^+(m_j) = \sum_{n=1}^{N} \frac{ m_j(n)~ PWM_{\textbf{j}}^+ (n)}{ \sum_{n=1}^{N}  PWM_{\textbf{j}}^+(n)}
\end{equation}

and 

\begin{equation}\label{mfPWMn}
Score^-(m_j) = \sum_{n=1}^{N} \frac{  ( ~1-m_j(n) ~) ~PWM_{\textbf{j}}^- (n)}{ \sum_{n=1}^{N} PWM_{\textbf{j}}^-(n)}
\end{equation}

where $\textbf{j} \in \Omega_k $ is defined in Eq. \ref{XC}.

 \begin{table*}[h]

  \caption{Comparison of the feature-size of proposed feature generation methods}
 
\label{tbl:Table_comp_Feature} 
\centering
\resizebox{0.9\linewidth}{!}{
\renewcommand{\arraystretch}{1.8}
\begin{tabular}{|c|c|c|c|c|c|c|}
\hline
 \textbf{Method} & \textbf{Parameters} & \textbf{Classifier} & \textbf{Feature  size} & \textbf{Accuracy} & \textbf{Sensitivity} & \textbf{Specificity} \\ \hline
\textit{Raw data} & -- & -- & 2400 & 60.06 & 78.27 & 41.84  \\ \hline
%  &  &  &  &  & \textit{SCSA} & h=100, Nh=47,B\_,S\_ &  & \textbf{2} & 89.30 & 91.17 & 87.43 & 87.92 & 89.27 & 89.50 & 0.959 \\
 \textit{PWM} & M=8, $\mathbf{r}$=0.18 & SVM  & \textbf{2} & 92.10 & 91.10 & 93.10  \\  \hline
 \textit{mPWM} & \begin{tabular}[c]{@{}l@{}}  M=10, $\mathbf{r}$=0.15  \\ \textit{k}-mers:  $\mathbf{j}=\{1,2\}$ \end{tabular}  &  SVM & 220 & \textbf{98.23} & \textbf{98.07} & \textbf{98.39}   \\ \hline
\end{tabular}

}
\end{table*}

\begin{table*}[!h]

  \caption{Comparison of the proposed methods with existing methods reported recently in \cite{Khalid2017}}
 
\label{tbl:Table_comp_exist} 
\centering
\resizebox{0.9\linewidth}{!}{
\renewcommand{\arraystretch}{1.4}
\begin{tabular}{|c|c|c|c|c|c|}
\hline
\textbf{Method} & \textbf{\begin{tabular}[c]{@{}c@{}}\#\\ Subject \end{tabular}} & \textbf{Feature} & \textbf{Accuracy} & \textbf{Sensitivity} & \textbf{Specificity} \\ \hline
ICA \cite{Ossadtchi2004} & 4 & \begin{tabular}[c]{@{}c@{}}Independent Components \\ Analysis\end{tabular} & - & 86.91 & 81.19 \\ \hline
CSP-LDA \cite{Khalid2016} & 20 & \begin{tabular}[c]{@{}c@{}}Common Spatial Patterns and \\ Linear Discriminant Analysis\end{tabular} & - & 86.14 & 90.38 \\ \hline
AT-DTW  \cite{Khalid2017} & 30 & \begin{tabular}[c]{@{}c@{}}Amplitude Thresholding and\\  Dynamic Time Warping\end{tabular} & - & 92.45 & 95.81 \\ \hline
 SCSA-RF  \cite{chah2019SCSA}  & 16 & \begin{tabular}[c]{@{}c@{}}Semi-Classical Signal Analysis\\   Random Forest  classifer  \end{tabular} & 94.33 &  93.68 & 95.08  \\ \hline
% SCSA  \cite{Khalid2017}  & \multirow{3}{*}{8} & \begin{tabular}[c]{@{}c@{}}   Semi-Classical Signal Analysis\\   SVM classifer\end{tabular} & 89.29 & 91.16 & 87.42 \\ \cline{1-1} \cline{3-6} 
PWM-SVM  & \multirow{2}{*}{16} &  \begin{tabular}[c]{@{}c@{}}  Position Weight Matrix\\   SVM classifer\end{tabular} & 92.10 & 91.10 & 93.10 \\ \cline{1-1} \cline{3-6} 
mPWM-SVM  &  & \begin{tabular}[c]{@{}c@{}}motif-based Position Weight Matrix\\   SVM classifer\end{tabular}  & \textbf{98.22} & \textbf{98.06} & \textbf{98.38} \\ \hline
\end{tabular}

}
\end{table*}

\subsection{Classification models}

The input MEG raw data after the pre-procesing step has  1734 spiky samples and 1734 healthy samples from different MEG test sessions of eight healthy and eight epileptic patients.The quantified samples are used to  generate QuPWM features and then   fed to  different classification models. The Support Vector Machine (SVM) model  outperforms the other models in  5-folds cross-validation (CV) process.The performance  is measured using the average accuracy, sensitivity, and specificity.

\section{Results}

\subsection{Comparison of the PWM-based features }
Each of the utilized PWM approaches captures a different property of the spikes characteristic. In order to study how these features perform and compare their performances, the different PWM approaches  have been tested on the same dataset, eight healthy and eight epileptic subjects, using the same classifier which is the SVM. The obtained results are shown in Table \ref{tbl:Table_SVM_L_sub}. The results show that the performance of the PWM method can be improved significantly using the di-mer feature. However, the  mPWM method needs larger memory and longer time. This limitation can be addressed using parallelization programming and feature selection to select the most important motif to extract.

 \begin{figure}[!h]
      \centerline{\includegraphics[width=0.75\linewidth]{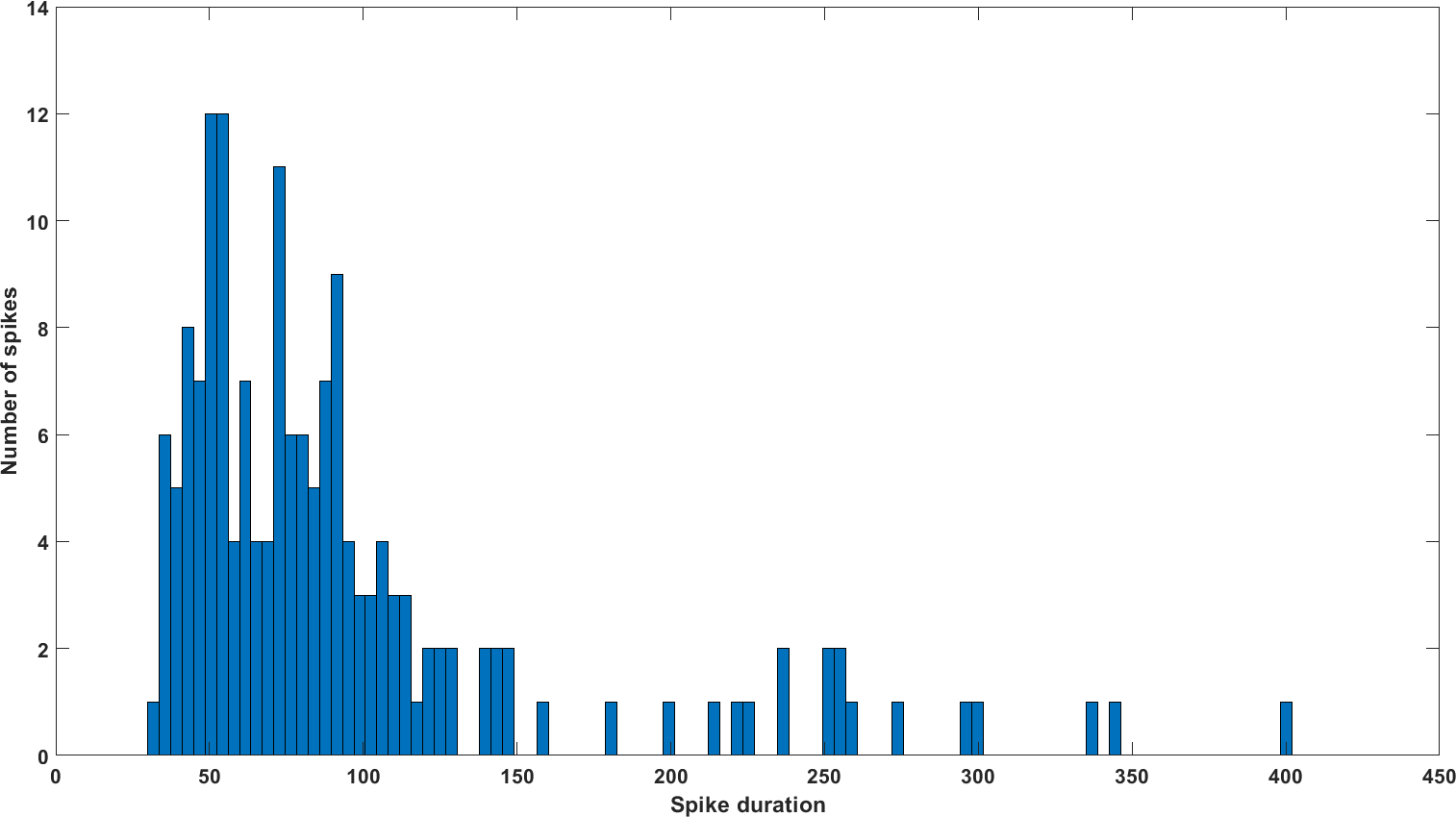}}
      \caption{The statistical properties of the epileptic spikes duration for eight epileptic subjects. }
      \label{fig:spikes_duration}
      \centering
  \end{figure}

\subsection{Comparison with existing methods}

 Several detection methods have been proposed for spikes detection using MEG signals.  The proposed features generation methods are compared to the reported performance of some recent works \cite{Khalid2017}. Table \ref{tbl:Table_comp_exist} shows the proposed feature improves the detection sensitivity and specificity especially using the mPWM features.  Table  \ref{tbl:Table_comp_Feature}  shows that the proposed methods have great potential in improving the detection accuracy and reducing the feature vector size.

\section{Discussion}

\begin{table*}[!t]
\centering
  \caption{ The 5-Folds cross-validation performance with  $L=70$ for different subjects combination.}
\label{tbl:Table_SVM_L_sub} 
\resizebox{\linewidth}{!}{
\renewcommand{\arraystretch}{1.3}

\begin{tabular}{|c|c|c|c|c|c|c|c|c|c|c|c|}
\hline
\textbf{\begin{tabular}[c]{@{}c@{}}\#\\ Subjects\end{tabular}} & \textbf{\begin{tabular}[c]{@{}c@{}}\#\\ Electrodes\end{tabular}} & \textbf{size} & \textbf{L} & \textbf{step} & \textbf{Method} & \textbf{Parameters} & \textbf{Classifier} & \textbf{Feature  size} & \textbf{Accuracy} & \textbf{Sensitivity} & \textbf{Specificity}  \\ \hline
\multirow{3}{*}{\textit{\textbf{16}}} & \multirow{3}{*}{24} & \multirow{3}{*}{4862} & \multirow{3}{*}{70} & \multirow{3}{*}{2} & \textit{Row\_Data} & -- & \multirow{3}{*}{SVM} & 1680 & 55.44 & 59.97 & 48.92  \\
%  &  &  &  &  & \textit{SCSA} &  &  &  &  &  &  &  &  &  &  \\
 &  &  &  &  & \textit{PWM} & M=8, $\mathbf{r}$=0.18 &  & 2 & 54.34 & 85.12 & 31.97   \\
 &  &  &  &  & \textit{mPWM} & M=12, $\mathbf{r}$=0.12 , $\mathbf{j}=\{1,2\}$&  & 312 & 66.98 & 81.52 & 52.43  \\ \hline
\multirow{3}{*}{\textit{\textbf{14}}} & \multirow{3}{*}{24} & \multirow{3}{*}{4730} & \multirow{3}{*}{70} & \multirow{3}{*}{2} & \textit{Row\_Data} & -- & \multirow{3}{*}{SVM} & 1680 & 55.48 & 63.13 & 47.82   \\
%  &  &  &  &  & \textit{SCSA} &  &  &  &  &  &  &  &  &  &  \\
 &  &  &  &  & \textit{PWM} & M=12, $\mathbf{r}$=0.12 &  & 2 & 58.54 & 85.12 & 31.97   \\
 &  &  &  &  & \textit{mPWM} & M=12, $\mathbf{r}$=0.12, $\mathbf{j}=\{1,2\}$ &  & 312 & 66.98 & 81.52 & 52.43   \\ \hline
\multirow{3}{*}{\textit{\textbf{12}}} & \multirow{3}{*}{24} & \multirow{3}{*}{4048} & \multirow{3}{*}{70} & \multirow{3}{*}{2} & \textit{Row\_Data} & -- & \multirow{3}{*}{SVM} & 1680 & 56.62 & 64.28 & 48.96   \\
%  &  &  &  &  & \textit{SCSA} &  &  &  &  &  &  &  &  &  &  \\
 &  &  &  &  & \textit{PWM} & M=8, $\mathbf{r}$=0.18 &  & 2 & 65.02 & 74.80 & 55.23   \\
 &  &  &  &  & \textit{mPWM} & M=12, $\mathbf{r}$=0.12, $\mathbf{j}=\{1,2\}$ &  & 312 & 68.36 & 77.62 & 59.09   \\ \hline
\multirow{3}{*}{\textit{\textbf{10}}} & \multirow{3}{*}{24} & \multirow{3}{*}{3464} & \multirow{3}{*}{70} & \multirow{3}{*}{2} & \textit{Row\_Data} & -- & \multirow{3}{*}{SVM} & 1680 & 59.12 & 70.95 & 47.28  \\
%  &  &  &  &  & \textit{SCSA} &  &  &  &  &  &  &  &  &  &  \\
 &  &  &  &  & \textit{PWM} & M=8, $\mathbf{r}$=0.18 &  & 2 & 64.72 & 72.40 & 57.05   \\
 &  &  &  &  & \textit{mPWM} & M=12, $\mathbf{r}$=0.12, $\mathbf{j}=\{1,2\}$ &  & 312 & 70.61 & 80.89 & 60.34   \\ \hline
\multirow{3}{*}{\textit{\textbf{8}}} & \multirow{3}{*}{24} & \multirow{3}{*}{3260} & \multirow{3}{*}{70} & \multirow{3}{*}{2} & \textit{Row\_Data} & -- & \multirow{3}{*}{SVM} & 1680 & 58.31 & 72.02 & 44.60   \\
%  &  &  &  &  & \textit{SCSA} &  &  &  &  &  &  &  &  &  &  \\
 &  &  &  &  & \textit{PWM} & M=10, $\mathbf{r}$=0.15 &  & 2 & 63.87 & 84.29 & 43.44   \\
 &  &  &  &  & \textit{mPWM} & M=10, $\mathbf{r}$=0.15, $\mathbf{j}=\{1,2\}$ &  & 220 & 72.15 & 84.05 & 60.25   \\ \hline
\multirow{3}{*}{\textit{\textbf{6}}} & \multirow{3}{*}{24} & \multirow{3}{*}{2728} & \multirow{3}{*}{70} & \multirow{3}{*}{2} & \textit{Row\_Data} & -- & \multirow{3}{*}{SVM} & 1680 & 60.67 & 74.56 & 46.78   \\
%  &  &  &  &  & \multicolumn{1}{l|}{\textit{SCSA}} & \multicolumn{1}{l|}{} &  &  &  &  &  &  &  &  &  \\
 &  &  &  &  & \multicolumn{1}{l|}{\textit{PWM}} & \multicolumn{1}{l|}{M=8, $\mathbf{r}$=0.18} &  & 2 & 64.66 & 81.00 & 48.30   \\
 &  &  &  &  & \multicolumn{1}{l|}{\textit{mPWM}} & \multicolumn{1}{l|}{M=12, $\mathbf{r}$=0.12, $\mathbf{j}=\{1,2\}$} &  & 312 & 73.46 & 87.90 & 59.02   \\ \hline
\multirow{3}{*}{\textit{\textbf{4}}} & \multirow{3}{*}{24} & \multirow{3}{*}{2644} & \multirow{3}{*}{70} & \multirow{3}{*}{2} & \textit{Row\_Data} & -- & \multirow{3}{*}{SVM} & 1680 & 60.10 & 76.63 & 43.57   \\
%  &  &  &  &  & \textit{SCSA} &  &  &  &  &  &  &  &  &  &  \\
 &  &  &  &  & \textit{PWM} & M=12, $\mathbf{r}$=0.12 &  & 2 & 94.82 & 94.70 & 94.93   \\
 &  &  &  &  & \textit{mPWM} & M=8, $\mathbf{r}$=0.18, $\mathbf{j}=\{1,2\}$ &  & 144 & 99.09 & 98.26 & 99.92   \\ \hline
\end{tabular}

}
\end{table*}

\begin{table*}[!t]
\centering

  \caption{The 5-Folds cross-validation performance with $L=100$ for different subjects combination}
\label{tbl:Table_SVM_L100_sub} 
\resizebox{\linewidth}{!}{
\renewcommand{\arraystretch}{1.3}

\begin{tabular}{|c|c|c|c|c|c|c|c|c|c|c|c| }
\hline
\textbf{\begin{tabular}[c]{@{}c@{}}\#\\ Subjects\end{tabular}} & \textbf{\begin{tabular}[c]{@{}c@{}}\#\\ Electrodes\end{tabular}} & \textbf{size} & \textbf{L} & \textbf{step} & \textbf{Method} & \textbf{Parameters} & \textbf{Classifier} & \textbf{Feature  size} & \textbf{Accuracy} & \textbf{Sensitivity} & \textbf{Specificity}   \\ \hline
\multirow{3}{*}{\textit{\textbf{16}}} & \multirow{3}{*}{24} & \multirow{3}{*}{3102} & \multirow{3}{*}{100} & \multirow{3}{*}{2} & \textit{Row\_Data} & -- & \multirow{3}{*}{SVM} & 2400 & 60.06 & 78.27 & 41.84   \\
 &  &  &  &  & \textit{PWM} & M=8, $\mathbf{r}$=0.18 &  & 2 & 92.10 & 91.10 & 93.10   \\
 &  &  &  &  & \textit{mPWM} & M=10, $\mathbf{r}$=0.15, $\mathbf{j}=\{1,2\}$ &  & 220 & 98.23 & 98.07 & 98.39   \\ \hline
\multirow{3}{*}{\textit{\textbf{14}}} & \multirow{3}{*}{24} & \multirow{3}{*}{3079} & \multirow{3}{*}{100} & \multirow{3}{*}{2} & \textit{Row\_Data} & -- & \multirow{3}{*}{SVM} & 2400 & 59.21 & 76.56 & 41.85   \\
%  &  &  &  &  & \textit{SCSA} & h=100, Nh=47,S\_ &  & 1 & 89.67 & 91.55 & 87.78   \\
 &  &  &  &  & \textit{PWM} & M=8, $\mathbf{r}$=0.18 &  & 2 & 92.76 & 92.40 & 93.11   \\
 &  &  &  &  & \textit{mPWM} & M=10, $\mathbf{r}$=0.15, $\mathbf{j}=\{1,2\}$ &  & 220 & 98.38 & 98.31 & 98.44   \\ \hline
\multirow{3}{*}{\textit{\textbf{12}}} & \multirow{3}{*}{24} & \multirow{3}{*}{2819} & \multirow{3}{*}{100} & \multirow{3}{*}{2} & \textit{Row\_Data} & -- & \multirow{3}{*}{SVM} & 2400 & 61.44 & 78.87 & 44.00   \\
%  &  &  &  &  & \textit{SCSA} & -- &  &  &  &  &  &  &  &  &  \\
 &  &  &  &  & \textit{PWM} & M=8, $\mathbf{r}$=0.18 &  & 2 & 91.98 & 90.92 & 93.04  \\
 &  &  &  &  & \textit{mPWM} & M=10, $\mathbf{r}$=0.15, $\mathbf{j}=\{1,2\}$ &  & 220 & 98.44 & 98.30 & 98.58   \\ \hline
\multirow{3}{*}{\textit{\textbf{10}}} & \multirow{3}{*}{24} & \multirow{3}{*}{2503} & \multirow{3}{*}{100} & \multirow{3}{*}{2} & \textit{Row\_Data} & -- & \multirow{3}{*}{SVM} & 2400 & 62.44 & 84.03 & 40.85   \\
%  &  &  &  &  & \textit{SCSA} & -- &  &  &  &  &  &  &  &  &  \\
 &  &  &  &  & \textit{PWM} & M=8, $\mathbf{r}$=0.18 &  & 2 & 94.17 & 94.49 & 93.84   \\
 &  &  &  &  & \textit{mPWM} & M=10, $\mathbf{r}$=0.15, $\mathbf{j}=\{1,2\}$ &  & 220 & 98.52 & 98.88 & 98.16   \\ \hline
\multirow{3}{*}{\textit{\textbf{8}}} & \multirow{3}{*}{24} & \multirow{3}{*}{2443} & \multirow{3}{*}{100} & \multirow{3}{*}{2} & \textit{Row\_Data} & -- & \multirow{3}{*}{SVM} & 2400 & 61.69 & 82.98 & 40.37   \\
%  &  &  &  &  & \textit{SCSA} & -- &  &  &  &  &  &  &  &  &  \\
 &  &  &  &  & \textit{PWM} & M=8, $\mathbf{r}$=0.18 &  & 2 & 95.13 & 94.85 & 95.41   \\
 &  &  &  &  & \textit{mPWM} & M=10, $\mathbf{r}$=0.15, $\mathbf{j}=\{1,2\}$ &  & 220 & 99.55 & 99.10 & 100.00   \\ \hline
\multirow{3}{*}{\textit{\textbf{6}}} & \multirow{3}{*}{24} & \multirow{3}{*}{2263} & \multirow{3}{*}{100} & \multirow{3}{*}{2} & \textit{Row\_Data} & -- & \multirow{3}{*}{SVM} & 2400 & 62.40 & 81.18 & 43.59   \\
%  &  &  &  &  & \multicolumn{1}{l|}{\textit{SCSA}} & \multicolumn{1}{l|}{--} &  &  &  &  &  &  &  &  &  \\
 &  &  &  &  & \multicolumn{1}{l|}{\textit{PWM}} & \multicolumn{1}{l|}{M=8, $\mathbf{r}$=0.18} &  & 2 & 95.27 & 95.14 & 95.40   \\
 &  &  &  &  & \multicolumn{1}{l|}{\textit{mPWM}} & \multicolumn{1}{l|}{M=10, $\mathbf{r}$=0.15, $\mathbf{j}=\{1,2\}$} &  & 220 & 99.51 & 99.03 & 100.00   \\ \hline
\multirow{3}{*}{\textit{\textbf{4}}} & \multirow{3}{*}{24} & \multirow{3}{*}{2246} & \multirow{3}{*}{100} & \multirow{3}{*}{2} & \textit{Row\_Data} & -- & \multirow{3}{*}{SVM} & 2400 & 63.00 & 82.19 & 43.81   \\
%  &  &  &  &  & \textit{SCSA} & -- &  &  &  &  &  &  &  &  &  \\
 &  &  &  &  & \textit{PWM} & M=8, $\mathbf{r}$=0.18 &  & 2 & 94.61 & 94.12 & 95.10   \\
 &  &  &  &  & \textit{mPWM} & M=10, $\mathbf{r}$=0.15, $\mathbf{j}=\{1,2\}$ &  & 220 & 98.04 & 96.97 & 99.11  \\ \hline
\end{tabular}

}
\end{table*}

\subsection{Choice of the frame length $L$}
As explained in Section \ref{Datata_frame}, the training and testing data are extracted from the MEG records using a sliding window or frame of size $L$.  The choice of the frame-size  is motivated by the statistical properties of the spikes time-duration for the different eight subjects. Most of the spikes last  almost $95$ sample-points$~= 95ms$ as shown in Figure \ref{fig:spikes_duration}. Therefore, the frame should be long enough to capture the common characteristics of the different spikes. In order to study  how sensitive are the proposed feature generation methods to the input datasets, we performed a sensitivity analysis to  study the effect of the frame-length and step-size on the obtained average performance in 5-folds cross-validation using SVM classifier with two frame-lengths $L=\{100,70\}$. Tables \ref{tbl:Table_SVM_L_sub}  and \ref{tbl:Table_SVM_L100_sub} demonstrate that a frame-length $100$ sample-points$= 100ms$  gives better detection accuracy.  Therefore, the optimal frame-length  should be around the mean spike duration for all subjects.

% \begin{enumerate}
% \item Different Frame lengths: $L=100$ and  $L=70$ 
% \item SVM  classifiers as it gives better results than the LR.
% \end{enumerate}

\subsection{Subject-independent classification}
For epileptic spikes detection, it is important to build subject-independent models which can provide higher specificity and sensitivity for any number of subjects. In order to study the effect of the number of subjects on the detection performance, we studied the combination of randomly combined subjects using SVM classifier and two frame-lengths $L=\{100,70\}$. The QuPWM-based features show a constant performance for most of the combinations using the optimal frame-size 100 sample-points, as shown in Table  \ref{tbl:Table_SVM_L100_sub} . This is because the QuPWM-based features extract the common patterns within the dataset regardless of its size.
% However, for a very big size data, this remark might not hold as shown in Tables \ref{tbl:Table_SVM_L_sub}  and \ref{tbl:Table_SVM_L100_sub}.

\section{Conclusions}

We developed a feature extraction method, called QuPWM, for epileptic spikes detection in MEG signals.  This method is based on combining the position weight matrix (PWM) method with digital quantization. The method shows great potential in improving the spike detection accuracy. Moreover, we achieved up to 98.06\% in sensitivity and 98.38\% in specificity for a dataset consisting of eight healthy and eight epileptic patients. A cluster of the input sequences can be adopted in the future, to build PWM-specific feature for each cluster of similar subjects, to build a hybrid detection model which might improve the accuracy, especially for outliers samples.

%%%%%%%%%%%%%%%%%%%%%%%%%%%%%%%%%%%%%%%%%%%%%%%%%%%%%%%%%%%%%%%%%%%%%%%%%%%%%%%%%%%%%
%
%     please remove the " % " symbol from \centerline{\includegraphics{fig01.eps}}
%     as it may ignore the figures.
%
%%%%%%%%%%%%%%%%%%%%%%%%%%%%%%%%%%%%%%%%%%%%%%%%%%%%%%%%%%%%%%%%%%%%%%%%%%%%%%%%%%%%%%

% \appendices

% % you can choose not to have a title for an appendix
% % if you want by leaving the argument blank
% \section{}

\section*{Acknowledgment}
 \noindent Research reported in this publication was supported by King Abdullah University of Science and Technology (KAUST) in collaboration with  King Abdulaziz City for Science and Technology (KACST)  and King Saud University (KSU). 
 
\section*{Funding}

\noindent This research project has been funded by King Abdullah University of Science and Technology (KAUST) Base Research Fund (BAS/1/1627-01-01), in collaboration with King Abdulaziz City for Science and Technology (KACST) and King Saud University (KSU).

% -------------------------------------------------------------------------
\bibliographystyle{unsrt}  
\bibliography{All_References}

\begin{thebibliography}{10}

\bibitem{Epidef}
Robert~S. Fisher, Walter van~Emde Boas, Warren Blume, Christian Elger, Pierre
  Genton, Phillip Lee, and Jerome Engel~Jr.
\newblock Epileptic seizures and epilepsy: Definitions proposed by the
  international league against epilepsy (ilae) and the international bureau for
  epilepsy (ibe).
\newblock {\em Epilepsia}, 46(4):470--472, 2005.

\bibitem{Hamalainen1993}
Matti H{\"{a}}m{\"{a}}l{\"{a}}inen, Riitta Hari, Risto~J Ilmoniemi, Jukka
  Knuutila, and Olli~V Lounasmaa.
\newblock {Magnetoencephalography---theory, instrumentation, and applications
  to noninvasive studies of the working human brain}.
\newblock {\em Reviews of Modern Physics}, 65(2):413--497, apr 1993.

\bibitem{Stefan2017a}
Hermann Stefan and Eugen Trinka.
\newblock {Magnetoencephalography (MEG): Past, current and future perspectives
  for improved differentiation and treatment of epilepsies}.
\newblock {\em Seizure}, 44:121--124, jan 2017.

\bibitem{Englot2015}
Dario~J Englot, Srikantan~S Nagarajan, Brandon~S Imber, Kunal~P Raygor,
  Susanne~M Honma, Danielle Mizuiri, Mary Mantle, Robert~C Knowlton, Heidi~E
  Kirsch, and Edward~F Chang.
\newblock {Epileptogenic zone localization using magnetoencephalography
  predicts seizure freedom in epilepsy surgery}.
\newblock {\em Epilepsia}, 56(6):949--958, jun 2015.

\bibitem{Baillet2017}
Sylvain Baillet.
\newblock {Magnetoencephalography for brain electrophysiology and imaging}.
\newblock {\em Nature Neuroscience}, 20:327, feb 2017.

\bibitem{El-Samie2018}
Fathi~E. {Abd El-Samie}, Turky~N. Alotaiby, Muhammad~Imran Khalid, Saleh~A.
  Alshebeili, and Saeed~A. Aldosari.
\newblock {A Review of EEG and MEG Epileptic Spike Detection Algorithms}.
\newblock {\em IEEE Access}, 6:60673--60688, 2018.

\bibitem{Ossadtchi2004}
A~Ossadtchi, S~Baillet, J~C Mosher, D~Thyerlei, W~Sutherling, and R~M Leahy.
\newblock {Automated interictal spike detection and source localization in
  magnetoencephalography using independent components analysis and
  spatio-temporal clustering}.
\newblock {\em Clinical Neurophysiology}, 115(3):508--522, 2004.

\bibitem{Khalid2016}
M~I Khalid, T~Alotaiby, S~A Aldosari, S~A Alshebeili, M~H Al-Hameed, F~S~Y
  Almohammed, and T~S Alotaibi.
\newblock {Epileptic MEG Spikes Detection Using Common Spatial Patterns and
  Linear Discriminant Analysis}.
\newblock {\em IEEE Access}, 4:4629--4634, 2016.

\bibitem{Khalid2017}
M~I Khalid, T~N Alotaiby, S~A Aldosari, S~A Alshebeili, M~H Alhameed, and
  V~Poghosyan.
\newblock {Epileptic MEG Spikes Detection Using Amplitude Thresholding and
  Dynamic Time Warping}.
\newblock {\em IEEE Access}, 5:11658--11667, 2017.

\bibitem{chah2019SCSA}
Abderrazak Chahid, Turky~N. Alotaiby, Saleh~A. Alshebeili, and Taous-Meriem
  Laleg-Kirati.
\newblock {Feature Generation and Dimensionality Reduction using the Discrete
  Spectrum of the Schr{\"{o}}dinger Operator for Epileptic Spikes Detection}.
\newblock In {\em 2019 41st Annual International Conference of the IEEE
  Engineering in Medicine and Biology Society (EMBC)}, 2019.

\bibitem{Laleg-Kirati2013}
Taous-Meriem Laleg-Kirati, Emmanuelle Cr{\'e}peau, and Michel Sorine.
\newblock Semi-classical signal analysis.
\newblock {\em Mathematics of Control, Signals, and Systems}, 25(1):37--61,
  2013.

\bibitem{brain_2018}
Todd B.~Bates.
\newblock {\em Human brain 3D model– stock image}.
\newblock Apr 2012.

\bibitem{bates_2018}
Alexmit.
\newblock {\em Epileptic seizures and depression may share a common genetic
  cause, study suggests}.
\newblock Jan 2018.

\bibitem{Taulu_2006}
S~Taulu and J~Simola.
\newblock Spatiotemporal signal space separation method for rejecting nearby
  interference in {MEG} measurements.
\newblock {\em Physics in Medicine and Biology}, 51(7):1759--1768, mar 2006.

\bibitem{gray1998quantization}
Robert~M. Gray and David~L. Neuhoff.
\newblock Quantization.
\newblock {\em IEEE Transactions on Information Theory}, pages 2325--2383,
  1998.

\bibitem{pollard1982quantization}
David Pollard.
\newblock Quantization and the method of k-means.
\newblock {\em IEEE Transactions on Information theory}, 28(2):199--205, 1982.

\bibitem{nikulin2011three}
MS~Nikulin.
\newblock Three-sigma rule.
\newblock {\em Encyclopedia of Mathematics, available at: http://www.
  encyclopediaofmath. org/index. php}, 2011.

\bibitem{3Sigma2011}
Friedrich Pukelsheim.
\newblock The three sigma rule.
\newblock {\em The American Statistician}, 48(2):88--91, 1994.

\bibitem{stormo1982use}
Gary~D Stormo, Thomas~D Schneider, Larry Gold, and Andrzej Ehrenfeucht.
\newblock Use of the ‘perceptron’algorithm to distinguish translational
  initiation sites in e. coli.
\newblock {\em Nucleic acids research}, 10(9):2997--3011, 1982.

\bibitem{hertz1999identifying}
Gerald~Z Hertz and Gary~D. Stormo.
\newblock Identifying dna and protein patterns with statistically significant
  alignments of multiple sequences.
\newblock {\em Bioinformatics (Oxford, England)}, 15(7):563--577, 1999.

\bibitem{staden1984computer}
Rodger Staden.
\newblock Computer methods to locate signals in nucleic acid sequences.
\newblock 1984.

\bibitem{Akhtar2010}
Malik~Nadeem Akhtar, Syed~Abbas Bukhari, Zeeshan Fazal, Raheel Qamar, and
  Ilham~A Shahmuradov.
\newblock {POLYAR, a new computer program for prediction of poly (A) sites in
  human sequences}.
\newblock {\em BMC genomics}, 11(1):646, 2010.

\bibitem{Tabaska1999}
Jack~E Tabaska and Michael~Q Zhang.
\newblock {Detection of polyadenylation signals in human DNA sequences}.
\newblock {\em Gene}, 231(1):77--86, 1999.

\bibitem{xia2012position}
Abderrazak Chahid, Turky~N. Alotaiby, and Laleg-Kirati Alshebeili, Saleh A.
  Taous-Meriem.
\newblock Position weight matrix, gibbs sampler, and the associated
  significance tests in motif characterization and prediction.
\newblock {\em Scientifica}, 2012, 2012.

\end{thebibliography}

\end{document}